\documentstyle[epsfig]{article}
\input{epsf}
\def\pin{$\pi N \;$}
\begin{document}
\thispagestyle{empty}
{\small \hfill KFA-IKP(TH)-1997-21}

\hfill November 1997\\

\bigskip

\begin{center}
{\Large \bf Working group on $\pi\pi$ and $\pi$N interactions}

\bigskip

{\Large Summary }

\medskip\bigskip\bigskip

Ulf-G. Mei\ss ner$^1$ (Convenor), Martin Sevior$^2$ (Convenor),
A. Badertscher$^3$,\\ 
B. Borasoy$^4$, P. B\"uttiker$^5$, G. H\"ohler$^6$, M. Knecht$^7$,
O. Krehl$^1$, J. Lowe$^8$,\\
M. Moj\v zi\v s$^{9}$, G. M\"uller$^{10}$, O. Patarakin$^{11}$,
M. Pavan$^{12}$, A. Rusetsky$^{13}$,\\
M.E. Sainio$^{14}$,
J. Schacher$^{15}$, G. Smith$^{16}$, S. Steininger$^1$, V. Vereshagin$^{17}$

\bigskip\bigskip

$^1$ FZ J\"ulich, Institut f\"ur Kernphysik (Th), D-52425 J\"ulich, Germany\\
$^2$ School of Physics, University of Melbourne, Parkville 3052,
Australia\\
$^3$ Institute for Particle Physics, ETH Z\"urich, Switzerland\\  
$^4$ Department of Physics and Astronomy, University of Massachusetts,
Amherst, MA 01003, USA\\
$^5$ Institut f\"ur Theoretische Physik, Universit\"at Bern, CH-3012
Bern, Switzerland\\
$^6$ Institut f\"ur Theoretische Teilchenphysik, University of Karlsruhe \\
     Postfach 6980 D-76128 Karlsruhe, Germany\\
$^7$ CPT, CNRS-Luminy, Case 907, F-13288 Marseille Cedex 9, France\\
$^8$ University of New Mexico, NM, USA, for the E865 collaboration\\
$^{9}$ Dept. Theor. Phys. Comenius University, Mlynska dolina SK-84215 Bratislava,
Slovakia\\
$^{10}$ Univ. Bonn, Institut f\"ur Theoretische Kernphysik, D-53115 Bonn, Germany\\
$^{11}$ Kurchatov Institute, Moscow 123182, Russia
for the CHAOS collaboration\\
$^{12}$ Lab for Nuclear Science, M.I.T., 77 Massachusetts Ave.,\\ Cambridge,
  MA 02139, USA\\
$^{13}$ Bogoliubov Laboratory of Theoretical Physics, JINR, 141980, Dubna, Russia\\
$^{14}$ Dept. of Physics, Univ. of Helsinki, P.O. Box 9, FIN-00014 Helsinki, Finland\\
$^{15}$ Universit\"at Bern, Laboratorium f\"ur Hochenergiephysik, Sidlerstrasse 5,\\
CH-3012 Bern, Switzerland\\
$^{16}$ TRIUMF, 4004 Wesbrook Mall, Vancouver,
 BC, Canada V6T 2A3\\
$^{17}$ Institute of Physics, St.-Petersburg State University,\\
St.-Petersburg, 198904, Russia\\  

\end{center}

\medskip
 
\noindent {\bf Abstract.} This is the summary of the working group on
$\pi\pi$ and $\pi$N interactions of the Chiral Dynamics Workshop in
Mainz, September 1-5, 1997. Each talk is represented by an extended
one page abstract. Some additional remarks by the convenors are added.

\vfill \eject

\noindent{\large {\bf Summary of the convenors}}

\bigskip

\noindent Here, we briefly summarize the salient results of the
talks and intense discussions in the working group. More details
are given in the one page summaries provided by each speaker.
We have ordered these contributions in blocks pertaining to theoretical
and experimental developments in the $\pi\pi$ and the $\pi$N
systems. To obtain a more detailed view of the present status, the
reader should consult the references given at the end of most of
the contributions.

\medskip
 
\noindent\underline{$\pi \pi:$} Both in standard and generalized CHPT,
two loop calculations for $\pi \pi$ scattering have been performed.
These have reached a very high precision which needs to be matched by
the $K_{\ell 4}$ data expected from DA$\Phi$NE and BNL and the pionium
measurement at CERN. The outstanding theoretical challenges are
twofold: First, a more detailed investigation of electromagnetic
corrections is mandatory. First steps in his direction have been
taken but the hard problem of analyzing the processes $K \to
\pi\pi\ell\nu_\ell$ and $\pi N \to \pi \pi N $ needs to be tackled.
 Second, the corrections to the Deser formula needed to evaluate
the scattering length difference $|a_0^0 - a_0^2|^2$ from the pionium
lifetime have to worked out precisely. Again, this problem is under
investigation and should be finished before the data will be analyzed.
At present, some discrepancies between the results of various groups
exist and these need to be eliminated.

\medskip
 
\noindent\underline{$\pi$N:} There has been considerable activity to
investigate elastic pion--nucleon scattering and the $\sigma$--term
in heavy baryon CHPT. The consensus is that these calculations have to
be carried out to order $q^4$ in the chiral expansion. For that, the
complete effective Lagrangian has to be constructed. This is under way.
Again, the remaining 
theoretical challenges are twofold: The em coorections need to be
looked at sytematically, for first steps see~[1]. Second, the
connection to dispersion theory has to be considered in more detail to
construct a more precise low--energy $\pi$N amplitude. Furthermore,
the program of partial wave analysis has to be refined to provide the
chiral community with precise input data like e.g. the pion--nucleon
coupling constant. Work along these lines is underway.

\medskip
\noindent In summary, considerable progress has been made since
the MIT workshop in 1994 and we are hopeful that this trend
continues until the next chiral dynamics workshop in the year 2000.

\medskip
\noindent [1] Ulf-G. Mei{\ss}ner and S. Steininger, [hep-ph/9709453], to appear in 
Phys. Lett. B.

\pagebreak

\noindent
{\Large\bf Low Energy $\pi-\pi$ Scattering to Two Loops in Generalized ChPT}

\bigskip

\noindent

Marc Knecht\\

CPT, CNRS-Luminy, Case 907, F-13288 Marseille Cedex 9, France

\bigskip

\noindent

The low energy $\pi-\pi$ amplitude $A(s\vert t,u)$ has been obtained to two 
loop accuracy in Ref. [1]. At this order, it 
is entirely determined by analyticity, crossing symmetry, unitarity and the 
Goldstone-boson nature of the [2], up to 
six independent parameters $\alpha$, $\beta$, $\lambda_{1,2,3,4}$, which are 
not fixed by chiral symmetry. Using existing medium energy $\pi-\pi$ 
data from unpolarized $\pi N\to\pi\pi N$ experiments, 
the values of the constants $\lambda_{1,2,3,4}$ were
 fixed from four rapidly convergent sum rules in [3].

Within the framework of Generalized ChPT [2], $\alpha$  can be related to 
the quark mass ratio $m_s/{\widehat m}$, or to the condensate 
$<{\bar q}q>_0$ through the ratio
 $x_{\rm GOR}=2{\widehat m<{\bar q}q>_0/F_{\pi}^2M_{\pi}^2}$. At leading 
order, $\alpha$ varies between 
1 ($x_{\rm GOR}=1$, $m_s/{\widehat m}=25.9$, the standard case), and 
4 ($x_{\rm GOR}=0$, $m_s/{\widehat m}=6.3$ , the extreme case of a 
vanishing condensate). 
The two loop expression of $A(s\vert t,u)$ in the standard case [4], 
together with the determination of $\lambda_{1,2,3,4}$ refered to above, 
lead to $\alpha=1.07$, $\beta=1.11$, corresponding to $a_0^0=0.209$ 
and $a_0^2=-0.044$ [5]. Therefore, a substantial departure of 
$\alpha$ from unity signals a much smaller value of the condensate than 
usually expected. A fit to the data of the Geneva-Saclay $K_{\ell 4}$ 
experiment [6] yields $\alpha=2.16\pm0.86$ and $\beta=1.074\pm0.053$. 
When converted into S, P, D and F-wave threshold parameters, these values 
reproduce the numbers and error bars obtained from the Roy equation analyses 
of the same data, {\it e.g.} $a_0^0=0.26\pm 0.05$, 
$a_0^2=-0.028\pm 0.012$ [7].
Given the values of $\lambda_{1,2,3,4}$ determined in [3], the two loop 
amplitude $A(s\vert t,u)$ of [1] thus becomes a faithful 
{\it analytic} representation of the numerical solution of the Roy equations 
from threshold up to $\sim 450$ MeV, where it satisfies the unitarity 
constraints. As a further example [3], the S- and P-wave phase shifts 
obtained from the two loop amplitude $A(s\vert t,u)$ are identical, in this 
energy range, to the numerical solution of the Roy equations for 
$a_0^0=0.30$ and $a_0^2=-0.018$ ({\it i.e.} 
$\alpha=2.84$ and $\beta=1.09$) quoted in [8].
In view of the theoretical implications, forthcoming and hopefully more 
precise low energy $\pi-\pi$ scattering data from the DIRAC experiment at 
CERN and from the $K_{\ell 4}$ experiments E865 at Brookhaven  and KLOE at 
DAPHNE are of particular interest and importance.

\noindent [1] M. Knecht et al., Nucl. Phys. B 457 (1995) 513 \\
\noindent [2] J. Stern et al., Phys. Rev. D 47 (1993) 3814. \\
\noindent [3] M. Knecht et al., Nucl. Phys. B 471 (1996) 445.\\
\noindent [4] J. Bijnens et al., Phys. Lett. B 374 (1996) 210; {\it ibid.}, hep-
ph/9707291\\
\noindent [5] L. Girlanda et al., hep-ph/9703448. To appear in Phys. Lett. B. \\
\noindent [6] L. Rosselet et al., Phys. Rev. D 15 (1977) 574 \\
\noindent [7] M.M. Nagels, Nucl. Phys. B 147 (1979) 189\\
\noindent [8] C.D. Froggatt and J.L. Petersen, Nucl. Phys. B 129 (1977) 89

\pagebreak

\noindent
{\Large \bf Roy Equation Studies of \boldmath{$\pi \pi$} Scattering} 

\bigskip

\noindent
B. Ananthanarayan,
Centre for Theoretical Studies, Indian Institute of Science,\\
Bangalore 560 012, India, and\\
\underline{P. B\"uttiker},
Institut f\"ur Theoretische Physik, Universit\"at Bern, CH-3012 Bern

\bigskip

\noindent $\pi \pi$ scattering is an important process to test the 
predictions of ChPT.
The $\pi\pi$ amplitude in its chiral expansion can be written as 
$A(s,t,u)=A^{(2)}(s,t,u) + A^{(4)}(s,t,u) + A^{(6)}(s,t,u) + O(p^8)$, where 
$A^{(2)}(s,t,u)$ is the Weinberg
result [1] and $A^{(4)}(s,t,u)$ and $A^{(6)}(s,t,u)$ are the one- 
and two-loop
contributions, respectively [2,3]. At leading order $a^I_0,
b^I_0$, and $a^1_1$ are the only non-vanishing threshold parameters 
($I=0, 2$), while
the one-loop (two-loop) calculation yields reliable predictions for five 
(eleven) more
threshold parameters.

$\pi\pi$ scattering has been studied in great detail in axiomatic
field theory. From dispersion relations, the Roy equations, a set of 
coupled integral
equations for the $\pi\pi$ partial wave amplitudes [4], have been 
derived. These
equations are used to derive sum rules for all the threshold parameters 
mentioned above
(exception: $a^0_0$ and $a^2_0$ which are the subtraction constants of the Roy
equations). Phase shift information, analyzed subject to respecting the 
Roy equations,
may then be used to evaluate the threshold parameters of interest.

We estimated the quantities of the higher threshold parameters for which no
information is available in the literature, using a modified effective range 
formula to
model the phase shift information, and
compared them, whenever possible, with the predictions of ChPT [5], 
assuming that
$a^0_0$ lies in the range favored by standard ChPT, i.e. $a^0_0\approx 0.21$. 
This comparison may be regarded as a probe into the range of validity in 
energy of chiral
predictions. Indeed, all the threshold parameters calculated in ChPT 
approach the values
calculated by sum rules when turning from the one-loop to the two-loop
calculation. We found that all except one higher threshold parameters in 
the two-loop
calculation (standard as well as generalized ChPT) are in good agreement
with the ones evaluated in the dispersive framework.
\\
\\

\noindent [1] S.~Weinberg, Phys.~Rev.~Lett.~17 (1966) 616.

\noindent [2] J.~Gasser and H.~Leutwyler, Ann.~Phys. 158 (1984) 142.

\noindent [3] J.~Bijnens et al., Phys.~Lett.~B~374 (1996) 210; 
hep-ph/9707291, M.~Knecht et al., Nucl.Phys. B 457 (1995) 513.

\noindent [4] S.~M.~Roy, Phys.~Lett.~B36 (1971) 353.

\noindent [5] B.~Ananthanarayan and P.~B\"uttiker, hep-ph/9707305, 
Phys.~Lett.~B, in print.

\pagebreak

\noindent
{\Large \bf Electromagnetic Corrections to $\pi\pi$--Scattering}

\bigskip

\noindent
Sven Steininger  \\
FZ J\"ulich, Institut f\"ur Kernphysik (Th), D-52425 J\"ulich, Germany

\bigskip

\noindent To perform a consistent treatment of isospin violation one has to
include virtual photons. Based on the observation that $\alpha=e^2/4\pi\simeq
M_{\pi}^2/(4\pi F_\pi)^2$, it is natural to assign a chiral dimension to the
electric charge, ${\cal O}(e) \sim {\cal O}(p)$ (see e.g. [1]). In [2]
the virtual photon Lagrangian is given up to fourth order:
\begin{eqnarray}
{\cal L}^\gamma_{\rm eff} = {\cal L}^\gamma_{\rm kin} + {\cal L}^\gamma_{\rm gauge}
+ {\cal L}^\gamma_2 + {\cal L}^\gamma_4 \qquad .
\end{eqnarray}
Theoretically the purest reaction to test the spontaneous and explicit chiral
symmetry breaking of QCD is elastic pion-pion scattering. In the threshold
region, the scattering amplitude in the isospin limit can be decomposed as 
$t^I_l = q^{2l} \left[ a^I_l + b^I_l q^2 + {\cal O}(q^4)\right]$, 
where $l$ denotes the pion angular momentum, $I$ the total isospin of the
two--pion system am $q$ the cms momentum. The S-wave scattering
lengths $a_0^{0,2}$ have been worked out to two loops in the chiral expansion
[3][4][5].
Including isospin violation one has to work in the physical basis, e.g. for the
process $\pi^0\pi^0 \rightarrow \pi^0\pi^0$ these have been worked out in
[2]. 
The result for the scattering length is $a_0(00;00)=0.034$ 
compared to 0.038 in the isospin limit. This decrease of 5\% comes entirely from the
correction of the Weinberg-term due to the pion mass difference and is
of the same size as the hadronic two loop contribution.  
A more dramatic effect appears in the effective range
$b_0(00;00)$=0.041 (0.030), which is related to the unitarity cusp
at $s=4M_{\pi^+}^2$. 

\noindent Since one is not able to measure pion-pion scattering directly, more 
involved
experiments have to be done to get experimental values. One of them, pion
induced pion production off the nucleon, has already been used in the isospin
symmetric case to pin down the isospin scattering lengths. Extending this to
the isospin violating sector one has to take into account as well the isospin
breaking in the $\pi N$--subsystem [6]. The most precise data at the
moment are given by $K_{l4}$ decays and experiments of this reaction at DA$\Phi$NE 
and BNL
should allow to examine isospin violation. In this case one has to calculate
the complete process of the decay including virtual photons. Since the photon
can couple to the charged lepton as well as to the charged mesons, an extension
of the effective field theory including virtual photons and leptons is needed. 
\vspace{-0.1cm}

\noindent [1] J. Gasser, in Proc. {\it Workshop on Physics and Detectors
    for DA$\Phi$NE'95}, Frascati Physics Series IV, 1995.\\
\noindent [2] Ulf-G. Mei\ss ner, G. M\"uller and S. Steininger, Phys. Lett. {\bf 
B406}, 154 (1997).\\
\noindent [3] S. Weinberg, Phys. Rev. Lett. {\bf 17}, 616 (1966).\\
\noindent [4] J. Gasser and H. Leutwyler, Phys. Lett. {\bf B125}, 325 (1983).\\
\noindent [5] J. Bijnens, G. Colangelo, G. Ecker, J. Gasser and M.E. Sainio,
    Phys. Lett. {\bf B374}, 210 (1996).\\
\noindent [6] Ulf-G. Mei\ss ner and S. Steininger, [hep-ph/9709453].\\

\pagebreak

\noindent
{\Large \bf Meson exchange models for $\pi\pi$ and $\pi\eta$ scattering}

\bigskip

\noindent
\underline{O. Krehl}, R. Rapp, G. Janssen, J. Wambach, and J. Speth  \\
FZ J\"ulich, Institut f\"ur Kernphysik (Th), D-52425 J\"ulich, Germany

\bigskip

\noindent We investigate the structure of $f_0(400-1200)(\sigma)$, $f_0(980)$
and $a_0(980)$ in the framework of coupled channel meson exchange
models for $\pi\pi/K\overline{K}$ and $\pi\eta/K\overline{K}$
scattering [1]. An effective meson Lagrangian is used to
construct a potential which contains $\rho$ exchange for $\pi\pi
\rightarrow \pi\pi$, $K^{\ast}(892)$ exchange for $\pi\pi,\pi\eta
\rightarrow K\overline{K}$ and $\rho,\omega,\phi$ exchange for
$K\overline{K} \rightarrow K\overline{K}$. Furthermore three contact
interactions - as appearing in the Weinberg-Lagrangian - and pole
graphs for $\rho$, $f_2(1270)$ and $\epsilon(1200-1400)$ formation are
included.

\noindent The such constructed potential is iterated within the
Blankenbecler Sugar equation.  This iteration is
necessary for the investigation of the nature of resonances, because
only the infinite sum will eventually lead to dynamical poles in
the scattering amplitude.  Due to the iteration each three meson
vertex has to be supplemented by a form-factor which parameterizes the
finite size of the three meson vertex. The cutoffs and coupling constants
(constrained by SU(3) symmetry relations)
are fixed by reproducing the experimental $\pi\pi$ data.

\noindent The $\pi\pi$ model leads to a very good description of
S,P and D-wave isoscalar and isotensor phase shifts up to 1.4GeV.  The
S-wave scattering lengths $a^0=0.210m_{\pi}^{-1}$ and
$a^2=-0.028m_{\pi}^{-1}$ are in good agreement with
experiment~[2][3][4] but $a^2$ slightly deviates from
two loop ChPT~[4]. The incorporation of minimal
chiral constraints~[2] i.e. including the $\pi\pi$ contact
terms and choosing an appropriate off shell prescription for the 0th
component of the 4-momenta ensures vanishing S-wave scattering lengths
$a^{I}$ in the chiral limit $m_{\pi} \rightarrow 0$.
By exploring the pole structure of the $\pi\pi$ amplitude we find a
broad $\sigma(400)$ pole at (II)$(468,\pm252)$MeV generated by $\rho$
exchange.  The sharp rise of the scalar isoscalar phase shift at
1.0GeV is produced by the narrow $f_0(980)$ $K\overline{K}$ bound
state pole on sheet (II) with $m_{f_0}=1005$MeV and
$\Gamma_{f_0}=50$MeV. At (III)$(1435,\pm181)$MeV we observe the
$\epsilon(1200-1400)$ pole, which we included in the potential as
effective description of higher scalar resonances or glueballs around
$1400$MeV.

\noindent For the $\pi\eta/K\overline{K}$ coupled channel we
find the $a_0(980)$ pole at (II)$(991,\pm101)$ MeV. This pole is
generated by the $\pi\eta\rightarrow K\overline{K}$ transition
potential and is therefore no bound state but a coupled channel pole.
Due to the nearby $K\overline{K}$ threshold the width
$\Gamma_{a_0}=202$MeV from the pole position is much
larger than the width $\Gamma_{a_0}\approx 110$MeV of the $a_0(980)$
resonance peak in the $\pi\eta$ cross section.

\noindent [1] G. Janssen {\it et al.}, Phys. Rev. D {\bf 52}, 2690 (1995).

\noindent [2] R. Rapp, J.W. Durso, J. Wambach, Nucl. Phys. {\bf A}596 (1996) 436.

\noindent [3] M.M. Nagels {\it et al.}, Nucl. Phys. {\bf B}147 (1979) 189.

\noindent [4] J. Bijnens {\it et al.} , Phys. Lett. {\bf B}374 (1996) 210.

\pagebreak

\noindent
{\Large \bf Hadronic Atoms as a Probe of Chiral Theory}

\bigskip

\noindent
V.E. Lyubovitskij and A.G. Rusetsky \\
Bogoliubov Laboratory of Theoretical Physics, JINR, 141980, Dubna, Russia

\bigskip

\noindent
At the present time DIRAC Collaboration at CERN is planning the
experiment on the high precision measurement of the $\pi^+\pi^-$
atom lifetime, which will provide direct determination of the
difference of the $S$-wave $\pi\pi$ scattering lengths $a_0-a_2$
with an accuracy 5~\% and thus might serve as a valuable test of
the predictions of Chiral Perturbation Theory [1].
In order to be able to compare the high-precision experimental
output with the theoretical predictions one has to carry out the
systematic study of various small corrections to the basic Deser-type
formula [2] which relates experimentally measured hadronic
atom lifetime to the $\pi\pi$ scattering lengths. This question has
been addressed in a number of recent publications
[3]-[6]. In papers [6] we present the
perturbative field-theoretical approach to the bound-state
characteristics, based on Bethe-Salpeter equation. It should be
emphasized that, in contrary with the nonrelativistic treatment of
the problem [3], we do not refer to the (phenomenological)
$\pi\pi$ interaction potential, which might introduce additional
ambiguity in the calculated observables of hadronic atoms. We achieve
a clear-cut separation of strong and electromagnetic interactions,
with all contribution from strong interactions concentrated in
$\pi\pi$ scattering lengths. In the Bethe-Salpeter framework we derive
the relativistic analogue of the Deser formula for the $\pi^+\pi^-$
atom decay width. The first-order corrections are
parametrized by the quantities $\delta_I\,\, (I=S, P, ...)$
$$
\Gamma=\frac{16\pi}{9}\sqrt{\frac{2\Delta m_\pi}{m_\pi}}
\sqrt{1-\frac{\Delta m_\pi}{2m_\pi}}
(a_0-a_2)^2|\psi_C(0)|^2(1+\delta_S+\delta_P+\delta_K+\delta_M+\delta_R)
$$
\noindent
We find that the sizeable contribution to Deser formula
$\delta_P=+1.85\%$ is due to the exchange of Coulombic photon
$t$-channel ladders and contains the nonanalytic ${\rm ln}\alpha$
term in the fine structure constant. We calculate also the corrections
coming from the displacement of the bound-state pole by strong
interactions  $\delta_S=-0.26\%$ and from the relativistic corrections
to the bound-state w.f. at the origin $\delta_K=-0.55\%$. The calculation
of the correction due to the $m_{\pi^\pm}-m_{\pi^0}$ mass difference
$\delta_M$ and the radiative correction $\delta_R$ is in progress.

\noindent [1]
J. Gasser and H. Leutwyler, Ann. Phys. (N.Y.) 158 (1984) 142.

\noindent [2]
S. Deser et al., Phys. Rev. 96 (1954) 774.

\noindent [3]
U. Moor, G. Rasche and W.S. Woolcock, Nucl. Phys. A 587 (1995) 747;\\
A. Gashi et al., nucl-th/9704017.

\noindent [4]
Z. Silagadze, JETP Lett. 60 (1994) 689;
E.A. Kuraev, hep-ph/9701327.

\noindent [5]
H.Jallouli and H.Sazdjian, hep-ph/9706450.

\noindent [6]
V. Lyubovitskij and A. Rusetsky, Phys. Lett. B 389 (1996) 181;\\
V.E. Lyubovitskij, E.Z. Lipartia and A.G. Rusetsky, hep-ph/9706244.

\pagebreak

\begin{center}
{\Large \bf Lifetime measurement of $\pi^+\pi^-$ atoms to test low energy QCD
predictions} \\

J\"urg\,Schacher \\
Universit\"at Bern,
Laboratorium f\"ur Hochenergiephysik, Sidlerstrasse 5,\\
CH-3012 Bern, Switzerland
\end{center}

The DIRAC Collaboration  
(DImeson Relativistic Atom Complex, PS212, CERN)
wants to measure the ground-state lifetime
of the exotic atom pionium, $A_{2\pi}$, formed by 
$\pi^+$ and $\pi^-$ mesons. There exists a relationship  
between this lifetime $\tau$ and the difference of
isoscalar minus isotensor s-wave scattering length: 
$\tau^{-1}=C\cdot \Delta^2$ with $\Delta=a_0-a_2$. 
If we aim to determine $\Delta$ down to 5\%,
corresponding to the theoretical uncertainty,
the lifetime has to be measured with 10\% accuracy - the 
very goal of the DIRAC experiment ([1]).

The method proposed takes
advantage of the Lorentz boost of relativistic pionium, produced in
high energy proton nucleus (e.g. Ti) reactions at 24 GeV/c (CERN
Proton Synchrotron). After production in the target, these
relativistic ($\gamma\simeq 15$) atoms may either decay into 
$\pi^0\pi^0$ or get excited
or ionized in the target material. In the case of ionization or 
breakup, characteristic charged pion pairs, called ``atomic pairs'',
will emerge, exhibiting low relative momentum in their centre of
mass system ($q< 3$MeV), small pair opening angle ($\theta_{+-}< 3$mrad
and nearly identical energies in the lab system 
($E_+\simeq E_-$ at the 0.3\% level).
The experimental setup is a magnetic double arm spectrometer 
to identify charged pions and to measure relative 
pair momenta $q$ with a resolution of $\delta q\simeq 1$MeV/c.
By these means, it is possible to determine the number of ``atomic pairs''
above background, arising from pion pairs in a free state. For a 
given target material and thickness, the ratio of the number of
``atomic pairs'' observed to the total amount of pionium produced depends on
the pionium lifetime $\tau$ in a unique way.

The experiment consists of coordinate detectors, 
a spectrometer magnet (bending power of 2 Tm) and two
telescope arms, each equipped with drift chambers, scintillation
hodoscopes, gas Cherenkov counters, preshower and muon detectors.
To reconstruct efficiently ``atomic pairs'' from pionium
breakup, the entire setup has to provide good charged particle
identification and extremely good relative cms momentum resolution. A 
high spatial resolution is guaranteed by the arrangement scintillating 
fibre detector - spectrometer magnet - drift chambers.
With a primary intensity of $1.5\cdot 10^{11}$ protons per spill 
(CERN PS) and a $\sim200\mu$m thick Ti target, a first level 
trigger rate of $4\cdot 10^{4}$ events per spill, due 
to free and accidental pairs, is expected. This high LHC-like trigger
rate will be reduced by a factor 30 by the trigger electronics.
Special purpose processors are intended to reject events 
with tracks more than 3 mrad apart as well as events with a large 
difference in the energies of the positively and negatively charged 
particles. The amount of data, necessary to measure $\tau$ 
with 10\% accuracy, should be collected in a running time 
of several weeks.

\noindent [1] 
B. Adeva  et al., {\it Lifetime measurement of $\pi^+\pi^-$ atoms to test low
energy QCD predictions}, Proposal to the SPSLC, CERN/SPSLC 95-1,
SPSLC/P 284, Geneva 1995.

\pagebreak

\begin{Large}
 
\centerline{\bf $K_{e4}$ measurements in Brookhaven E865}
 
\end{Large}
 
\vspace{0.1in}
 
\centerline{J. Lowe, University of New Mexico, for the E865 collaboration}
 
\vspace{0.15in}
 
\begin{figure}[h]
\epsfig{file=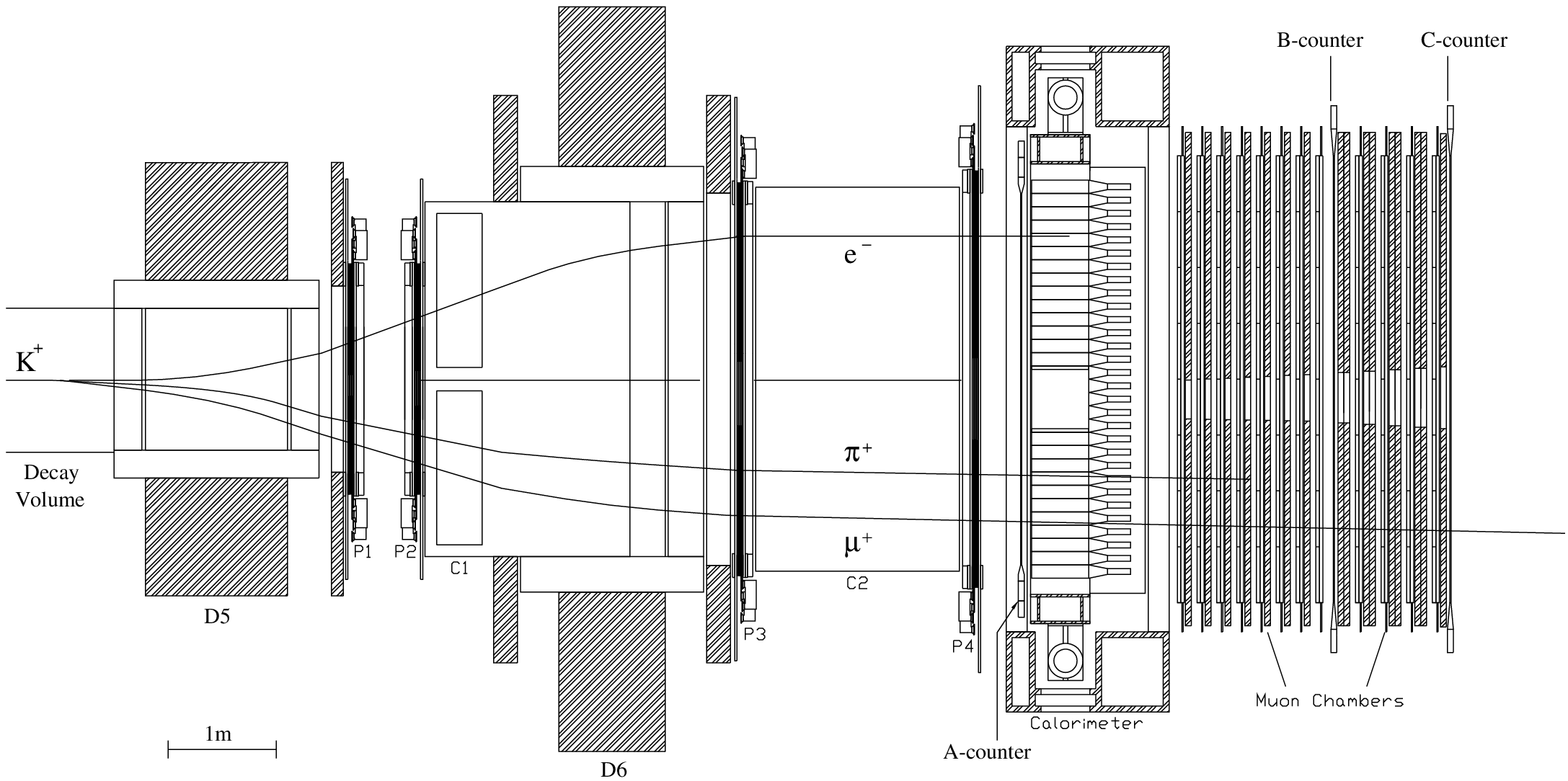,height=9.0cm, width=6.5cm,angle=90}
\end{figure}
 
The experiment consists of 6 GeV/c unseparated kaon beam,
a 5m Decay tank, a Magnet (D5) to separate positives and  negatives,
a Trigger hodoscope (A-counter), Momentum determination by PWCs
(P1 - P4) and magnet D6, Particle identification by an 
Electromagnetic shower calorimeter, {\v C}erenkov detectors (C1 and C2)
Muon range stack of iron, wire chambers and hodoscopes,
and a Beam tracker (not shown) upstream of  
the decay tank.

\noindent History and status of Brookhaven E865 
 
\vspace{0.05in}
 
\noindent 1993 - 5:  Development of beam line; detectors installed and 
commissioned; checked trigger rates, backgrounds, etc.
 
\noindent 1995 - 6:  Data taking on $K^+\rightarrow\pi^+\mu^+e^-$,
$K^+\rightarrow\pi^+e^+e^-$, etc. 
 
\noindent 1997: \hspace{0.18in} Data run on $K^+\rightarrow\pi^+\mu^+\mu^-$,
$K_{e3}$, $K_{e4}$; analysis just started.
 
\noindent (1998: \hspace{0.13in} Plan for long data run on 
$K^+\rightarrow\pi^+\mu^+e^-$)
 
\vspace{0.1in}
 
\noindent We have $\geq 3\times 10^5$ events on tape -- we hope to be
left with close to $3\times 10^5$ events after cuts and removal of
contaminant events.
 
\vspace{0.05in}
 
\noindent We plan to analyse the data in conjunction with the theoretical
group: Bijnens (Lund), Colangelo (Frascati), Ecker (Wien), Gasser (Bern), 
Knecht (Marseille), Meissner (J\"{u}lich), Sanio (Helsinki), Steininger 
(Bonn) and Stern (Orsay).
 

The phase space coverage in the $m_{e\nu}$ and $m_{\pi\pi}$ variables
is reasonably uniform, apart from at the highest $m_{\pi\pi}$. 

At this stage we are considering the following questions. 
How many events will we have after all cuts? 
We had problems with the beam tracker during the run and it may not be 
available for  all the data. How well can we analyse without it?
Is the phase-space coverage good enough?

\pagebreak

\begin{center}
{\Large \bf
 A measurement of the
$\pi^{\pm} p \rightarrow \pi^+ \pi^{\pm}n$ reactions Near Threshold }

O.~Patarakin,Kurchatov Institute, Moscow 123182, Russia \\
For the CHAOS collaboration.
\end{center}

 The pion induced pion production $\pi^{\pm} p \rightarrow \pi^+ \pi^{\pm}n$
reactions were studied at projectile incident kinetic energies (T$_{\pi}$)
of 
223, 243, 264, 284 and
305 MeV. The
CHAOS spectrometer (at TRIUMF) was used for the measurement.
Double differential cross sections were used as input to
the Chew-Low-Geoble
extrapolation procedure which was utilized to determinate on-shell
$\pi \pi$ elastic
scattering cross sections in the near threshold region.

  The pseudo-peripheral-approximation method was aplied, which
extrapolates an auxiliary function $F^{\prime} = F/|t|$ to the pion pole.
This method makes use of the fact that in the case of one-pion-exchange (OPE
) dominance, $F^{\prime} (t, m_{\pi \pi})$ is linear in $t$, which implies
$F(0, m_{\pi \pi}) = 0$. The Chew-Low-Geoble
 extrapolation must be performed under
conditions which enchance the OPE and suppress the background. This was
accomplished by carefully choosing the $t$-intervals over which the $F^
{\prime}(t, m_{\pi \pi})$ could be described by a linear function, and
required that the
condition $F(t=0, m_{\pi \pi})=0$ was satisfied. The we also
require that the COM $\pi \pi$~ angular distributions ($\cos{\Theta}$)
be flat.
 No attempt was made to remove potential background contributions
in this work, since there exists no reliable model for such a procedure.
Instead, we restrict our analysis to those regions where the
conditions discussed above are satisfied.
We followed the same Chew-Low-Geoble procedures as have been applied previously
at higher energies.

  For $\pi^+ \pi^-$ channel the extrapolation was performed for each
bin of $m_{\pi \pi}$. Only fits with $\chi^2 /\nu \le 1.8$ are used in the
analysis. We obtained cross section values for all
initial energies except 305 MeV.
At the largest values of  $m_{\pi \pi}$ and at T$_{\pi}$~ = 305 MeV, we
found that the form
for $F^{\prime}$ function was not linear but "bell-shaped".
It is very similar to the form for the $F^{\prime}$ function found,
if we take the $\Delta$-isobar peak events from the
experiment at 1.5 GeV/c. 
  The bulk of the cross section values obtained at different projectile 
energies agree within the error bars.

The resulting averaged cross sections are consistent with Roy equation
predictions based on previously measured $\pi \pi$ cross sections obtained
for the 5 charge
reaction channels at higher energies. This does not prove the validity of
the Chew-Low-Geoble
 technique as a tool for studying $\pi \pi$ scattering. However,
it suggests that this method can be applied in the same manner to both high
energy and threshold $\pi N \rightarrow \pi \pi N$ data.

The Roy equations were applied in order to obtain a self-consistent
determination of $\pi \pi$ scattering amplitudes. Taking into
account the present $\pi \pi$ cross sections,
the isospin zero S-wave scattering length is determined to be
$a^0_0 = 0.215 \pm 0.030$ in inverse pion mass units.

 It was also shown that for $\pi^+ \pi^+$ reaction, the Chew-Low-Geoble
 analysis was not
possible. The conditions for using the extrapolation technique, as
described above, were not satisfied for these data. The $\cos{\Theta}$
distributions were not flat, and the dependence of $F^{\prime}$ on $t$ had no
distinguishable linear region. 

\pagebreak

\begin{center}

{\Large \bf Determination of the
       \mbox{$\pi^{\pm}$ p $\rightarrow$ $\pi^{\pm}$ $\pi^{+}$ n}
       Cross-Section Near Threshold\\}

\medskip

M.E.Sevior   \\
{\em School of Physics,University of Melbourne,
                          Parkville Victoria, 3052, Australia}
\end{center}

One of the most fruitful ways of
investigating the  $\pi -\pi$ interaction
  interaction experimentally has involved the measurement
of threshold pion-induced pion production cross-sections.
The amplitudes for these reactions near threshold are dominated by the
One Pion Exchange process, which in turn can be related to the $\pi - \pi$
scattering process.
Bernard, Kaiser and Mei$\ss$ner~[1] used Baryon
Chiral Perturbation Theory, 
to predict 
amplitudes for pion production and to determine 
the $\pi-\pi$ scattering lengths.

The \mbox{$\pi^{-}$ p $\rightarrow$ $\pi^{-}$ $\pi^{+}$ n }
reaction involves both isospin 2 and isospin 0 $\pi \pi$ interaction
amplitudes and the \mbox{$\pi^{+}$ p $\rightarrow$ $\pi^{+}$ $\pi^{+}$ n }
reaction involves only isospin 2. 
We have employed the ``active target" system developed by Sevior
et al. ~[2][3] to measure  both processes and so have determined both
amplitudes near threshold.


The experiment was performed at TRIUMF at 200, 190, 184,
180 and 172 MeV for the negative pions and at 200, 184 and 172 MeV for the
positive.  
The cross sections measured by the experiment are summarized in Table 1.

\begin{table}[h]
 \begin{tabular}{c||c c | c||c}
          & \multicolumn{4}{c}{Cross sections ($\mu b$)} \\
 $T_{\pi}$ & \multicolumn{3}{c||}{$\pi^{+}p\rightarrow \pi^{+}\pi^{+}n$} &
             {$\pi^{-}p\rightarrow \pi^{-}\pi^{+}n$} \\
\cline{2-5}
 (MeV)     & One $\pi$     & Two $\pi$     & Averaged & One $\pi$ \\
 \hline
 200       & $1.4 \pm 0.3$ & $1.4 \pm 0.3$ & $1.4 \pm 0.3 $ & $6.5 \pm 0.9$ \\
 190       &       --      &       --      &       --       & $3.0 \pm 0.5$ \\
 184       & $.30 \pm .07$ & $.30 \pm .07$ & $.30 \pm .07 $ & $1.9 \pm 0.3$ \\
 180       &       --      &       --      &       --       & $0.7 \pm 0.1$ \\
 \end{tabular}
 \caption{Total cross-sections for
          \mbox{$\pi^{\pm}$ p $\rightarrow$ $\pi^{\pm}$ $\pi^{+}$ n}.  The
          uncertainties include both statistical and systematic errors.}
 \label{tab:sigma}
\end{table}

  Our cross section data yield threshold values for the
amplitudes : $\left| {\cal A}_{10} \right|  = (8.5 \pm 0.6)m_{\pi}^{-3}$
and $\left| {\cal A}_{32} \right| = (2.5 \pm 0.1)m_{\pi}^{-3}$,
and for the $\pi-\pi$ scattering
lengths: $a_{0} = ( 0.23  \pm 0.08 )m_{\pi}^{-1}$,
and $a_{2} = (-0.031 \pm 0.008)m_{\pi}^{-1}$.  
Our value for $\left| {\cal A}_{10} \right|$ is in good
agreement with the value
of $8.0  \pm 0.3 m_{\pi}^{-3}$ obtained by Bernard et al~[1]
and our values of the scattering lengths
are consistent with the Chiral Perturbation Theory
predictions of 
a$_{0} = (0.20 \pm 0.1) m_{\pi}^{-1}$~ and a$_{2} = (-0.042 \pm 0.02) 
m_{\pi}^{-1}$.
The uncertainties in
the extracted values of the scattering lengths are dominated by the
theoretical uncertainties.\\


\noindent [1] V.Bernard, N.Kaiser and Ulf-G.Mei$\ss$ner, Int. J. Mod. Phys
                {\bf E4} (1995) 193.;
                Nucl. Phys.
                {\bf B457} (1995) 147.;
                Nucl. Phys. {\bf A619} (1997) 261.

\noindent [2] M.E.Sevior {\em et al.}, Phys. Rev. Lett.
                {\bf 66} (1991) 2569.

\noindent [3] K.J.Raywood {\em et al.}, Nucl. Inst. and Meth.
                {\bf A365} (1995) 135.


\pagebreak 

\noindent
{\Large \bf Renormalization of the pion nucleon interaction to fourth order} 

\medskip

\noindent
Guido M\"uller \\
Univ. Bonn, Institut f\"ur Theoretische Kernphysik, D-53115 Bonn, Germany

\bigskip

\noindent We renormalize in the framework of HBCHPT 
the complete one--loop generating functional to order
q$^4$. In heavy baryon formalism the one-to-one correspondence between
the loop expansion and the chiral dimension is restored: one--loop
Feynman diagrams with only insertions of the lowest order meson baryon
Lagrangian can be renormalized by introducing the most general 
counterterm Lagrangian of dimension three and  
one--loop diagrams with
exact one insertion of the second order Lagrangian contribute to 
fourth order. This one-to-one correspondence allows to regularize
separately the one--loop generating functional to third and to fourth
order in the chiral dimension. The renormalization can be done by
separating the low energy constants into a finite and a divergent
part. The beta functions are chosen in such a way to cancel the
divergences of the one-loop generating functional. 
To third order
the beta functions depend in the two flavor case 
on the axial vector coupling constant $g_A$ [1].
To fourth order the beta
functions become functions of the low energy constants of the 
next-to-leading order Lagrangian of dimension two, 
i.e., $ g_A,m, c_1, \dots , c_7 $. The low energy constants which
appear in the effective Lagrangian of order one and two are always finite
and scale independent. 
The renormalization can be extended to the three flavor case [2].
The structure of the singular behavior is not changed
by this extension.
Since 
in SU(2) the nucleons are in the fundamental representation
and in SU(3) the baryons are in the adjoint representation, 
the evaluation of the divergences
is more complicated.
In the three flavor case the beta functions to third order depend on
the two axial vector couplings  $D$ and $F$.
The renormalization of the one--loop generating functional can be done 
by evaluating the path integral. 
To third order we find four typs of one--loop diagrams;
two reducible diagrams and two irreducible diagrams [1] [2].
The sum of the reducible diagrams is finite and the divergences
are given by the irreducible diagrams. The divergences can be 
extracted by using heat kernel methods for the meson and
baryon propagator. The most difficult part is to find a heat
kernel representation for the baryon propagator which is 
not elliptic and definite in Euclidian space [1]. 
To the fourth order we find four new irreducible diagrams
with exact one insertion of the second order Lagrangian.
In principle the renormalization can be done with the same
methods as for the renormalization to third order. The only
difference appears in one diagram where the dimension two 
insertion is on the intermediate baryon line. 
%
With the developed methods
one is able to evaluate
the three flavor case to fourth order, one can introduce virtual photons
to the strong sector. This is of interest for evaluating the
isospin violation effects in pion nucleon scattering or 
pion photoproduction. Another extension is the consideration
of nonleptonic or radiative hyperon decays. 

\smallskip

\noindent [1] G.Ecker, Phys. Letters {\bf B 336} (1994) 508 

\noindent [2] G.M\"uller and U.Mei\ss ner, Nucl. Phys. {\bf B 492} (1997) 379

\noindent
{\Large \bf $\pi N \;$ in Heavy Baryon ChPT} 

\bigskip

\noindent
Martin Moj\v zi\v s  \\
Dept. Theor. Phys. Comenius University, Mlynska dolina SK-84215 Bratislava,
Slovakia

\bigskip

\noindent Elastic \pin scattering amplitude was calculated up to the third
order in HBChPT [1]. The calculation was based on the Lagrangian [2],
containing 7 and 24 LECs (Low Energy Constants) at 2nd and 3rd order
respectively. Only a subset of 9 linear combinations of these LECs contributes
to the process at hand.

\noindent Comparison of the result with (extrapolated) experimental data was
done for a set of 16 threshold parameters [3], \pin $\sigma$-term and GT
discrepancy. 6 out of these 18 quantities do not depend on LECs of the 2nd and
3rd orders, and their comparison to the extrapolated experimental data seems
to be encouraging:
$$
\begin{tabular}{|l|l|l|l|l|l|l|} \hline
       & $\ a^+_{2+}$ & $\ a^-_{2+}$ & $\ a^+_{3+}$ & $\ a^+_{3-}$ & 
$\ a^-_{3+} $ & $\ a^-_{3-}$  \\ \hline
theory & $-36$        & $56$         & $280$        & $\ 31$       & $-210$     
  & $\ 57$ \\ \hline
data   & $-36\pm 7$   & $64\pm 3$    & $440\pm 140$ & $160\pm 120$ & 
$-260\pm 20 $ & $100\pm 20$\\ \hline
\end{tabular}
$$
\noindent where units for D-waves ($a_2$) are $GeV^{-5}$ and units for F-waves
($a_3$) are $GeV^{-7}$. 

\noindent It is instructive to see how do the separate orders of the chiral
expansion contribute to this result. However, one has to be a little bit
careful in what one calls a chiral order here. For two different possibilities
see [1] and [5].

\noindent Remaining 12 quantities dependent on LECs of 2nd and 3rd order can
be used for determination of these LECs. The result is:
$$
\begin{tabular}{|c|c|} \hline
$a_1$ & $-2.60 \pm 0.03$ \\ \hline
$a_2$ & $\ \ 1.40 \pm 0.05$ \\ \hline
$a_3$ & $-1.00 \pm 0.06$ \\ \hline
$a_5$ & $\ \ 3.30 \pm 0.05$ \\ \hline
\end{tabular}
\hskip 3cm
\begin{tabular}{|c|c|} \hline
$\tilde b_1 + \tilde b_2$ & $\ \ 2.4 \pm 0.3$ \\ \hline
$\tilde b_3$ & $-2.8 \pm 0.6$ \\ \hline
$\tilde b_6$ & $\ \ 1.4 \pm 0.3$ \\ \hline
$b_{16} - \tilde b_{15}$ & $\ \ 6.1 \pm 0.6$ \\ \hline
$b_{19}$ & $-2.4 \pm 0.4$ \\ \hline
\end{tabular}
$$
where $\tilde b_i$ is a renormalization-scale independent part of the
renormalization scale dependent quantity $b_i$ [1]. Values for 2nd order LECs
$a_i$ are in a quite good agreement with their recent determination in [4]. In
case of 3rd order LECs, this is their first (rough) determination.

\noindent Description of the considered set of data is not bad, but the results 
seem to strongly suggest calculation to the fourth order and comparison to a
larger set of data. For more details see [1] and [5].

\noindent [1] M.Moj\v zi\v s, hep-ph/9704415, to appear in Z.Phys. C

\noindent [2] G.Ecker and M.Moj\v zi\v s, Phys.Lett. B365 (1996) 312

\noindent [3] R.Koch and E.Pietarinen, Nucl.Phys. A336 (1980) 331

\noindent [4] V.Bernard, N.Kaiser and U.-G.Mei\ss ner, Nucl.Phys. A615 (1997) 483

\noindent [5] G.Ecker, these proceedings

\pagebreak

\begin{center}

\noindent{\large \bf Relations between Dispersion Theory \\
and Chiral Perturbation Theory}

\bigskip

\noindent
G. H\"ohler  \\
Institut f\"ur Theoretische Teilchenphysik, University of Karlsruhe \\
     Postfach 6980 D-76128 Karlsruhe, Germany
\end{center}

\bigskip
\noindent
The relations are given as comments to several sections in Ref. [1],
using the same notation.\\

\noindent
{\bf Calculation of the Isoscalar Spin-flip Amplitude}

Aside from a negligible contribution of the S-wave scattering length
$a^+_{0+}$, the isoscalar \pin flip amplitude $P^+_2$ is related to the
value of the invariant \pin amplitude $B^+$ at threshold by
\begin{equation}  
B^+ (th) = \frac{2m}{1+\mu} P^2_+ .
\end{equation}

The forward dispersion relation[2] has a pole term which agrees up to 
1\% with the Borntern in Ref.[1]. The dispersion integral contains
$P^2_+ (loop)$. $P^2_+ (\Delta)$ is interpreted as the $\Delta$-contribution
to the integral. Its numerical value lies between the crude estimates in 
Ref.[1]. A pole term approximation is given on p.563 in 
Ref.[2].
The unsolved problems with the $\Delta$-propagator and the $\pi N\Delta$ 
coupling 
constant do not occur.

The difference $a_{13} -a_{31}$ follows from the circle cut (t-channel 
exchanges) in the partial wave dispersion relation [3]. This term is
related to the loop.\\

\noindent
{\bf The S-wave Effective Range Parameter $b^-$}

From an exact projection of fixed-t dispersion relations[3], Koch
obtained an improved value 0.11 instead of 0.19 in Eq.(45) of 
Ref.[1]. In this calculation as well as in the combination with
Eq.(44), which occurs in Geffen's sum rule (p.281 in Ref.[2]), a
$\Delta$ contribution plays an important rule, which does not occur in
Ref.[1].\\

\noindent
{\bf Sect.6.3: D- and F-Wave Threshold Parameters}

There is a close relation to the fixed-s dispersion relation for s
taken at threshold[3].\\
A more detailed treatment of these and other topics will be available soon.

\noindent [1] V. Bernard, N. Kaiser, Ulf.-G. Mei{ss}ner:
  Nucl. Phys. A{\bf 615} 483 (1997)

\noindent [2] G. H\"ohler in Pion-Nucleon Scattering, Landoldt-B\"ornstein
I/9b2, ed. H. Schopper, Springer 1983

\noindent [3] R. Koch, Z. Physik C {\bf 29},597 (1985), Nucl. Phys. A {\bf 448}
707 (1986)

\pagebreak

\begin{centering}
{\Large \bf Recent Results from $\pi$N Scattering}\\
{\em Greg Smith, TRIUMF}\\
\end{centering}
\medskip

The primary physics issues to be addressed here are the determination
of $\Sigma_{\pi N}$, from which the strange sea quark contribution to
the proton wavefunction can be determined, and related issues such as
the $\pi$N and $\pi\pi$ scattering lengths, the $\pi$N coupling constant,
the values of the $\pi$N partial wave amplitudes (PWA), and the search
for signs of isospin violation.
Since the time of the previous Chiral Dynamics Workshop, there has
been a great deal of progress in $\pi$N scattering both experimentally
and theoretically. The TRIUMF $\pi$N experimental program has focussed
on measurements of $\pi^\pm p \longrightarrow \pi^\pm \pi^+ n$ at
energies near threshold, from which $\pi\pi$ scattering cross sections
can be derived as well as $\pi\pi$ scattering lengths. This topic is
discussed in the talk of Patarakin, and our result
for a$^0 _0$ of $0.215\pm 0.030$ is presented there.
Experiments even closer to threshold are discussed in the talk by
Sevior, and ($\pi,2\pi$) is reviewed by Pocanic.

Our previous work at TRIUMF concentrated on precise measurements of
$\pi^\pm$p differential cross sections. We present an excitation
function of the entire $d\sigma /d\Omega$ database at a couple of
representative angles. This clearly shows that the Bertin, et al. data
are outliers and that the rest of the database with a few minor
exceptions is in reasonable agreement with itself as well as with
SM95, but the KH80 PWA badly over-predicts the cross sections, an
effect which becomes more acute close to threshold.
Several new analyses
(Matsinos, Pavan, Timmermans, Gibbs) support the conclusions
with respect to the database and in fact use the low energy data to
examine isospin breaking. Both Matsinos and Gibbs report isospin
breaking effects at the 7\% level.

Our present efforts at TRIUMF are geared towards measurements of the
$\pi^\pm \vec{p}$ analyzing powers. We have recently completed
measurements at resonance energies which again show a clear preference
for the VPI PWA over the KH80 solution. Single-energy PWA has been
used to explore the sensitivity of the data to the values of the
S- and P-wave phase shifts. Measurements in the low energy
regime, in particular at the S-P interference minimum, are planned for
fall '97. These measurements have been shown to be especially
sensitive to the $\pi$N scattering lengths. A similar effort
at forward angles will be mounted at PSI, where use will be made of a
polarized scintillator target.

Our future $\pi$N program
consists of precise measurements of low energy $\pi^\pm$p
$d\sigma /d\Omega$
in the Coulomb-nuclear interference region. We have
shown how such data can provide a direct measure of $a^+ _{0+}$
and $a^+ _{1+}$.

The new experimental results clearly favor smaller values for f$^2 /4\pi$
than the canonical KH80
value of 0.079, in agreement with all recent
PWAs based on np, pp, N${\rm \bar{N}}$, and $\pi$N data.
The single exception is an analysis of Erickson and Loiseau, based on
an np scattering experiment at a single energy, which obtains a result
even higher than KH80. However, the consensus
from MENU97 was overwhelmingly in
favor of lowering the default value of f$^2 /4\pi$ to $\sim
0.075\pm0.01$.

A reliable determination of  $\Sigma_{\pi N}$~ requires
a careful sub-threshold
analysis ala KH80, but with the much more precise amplitudes
provided by either the VPI or Nijmegen analyses.

\pagebreak

\begin{center}
{\Large \bf Pion-Nucleon Scattering Lengths from Pionic Hydrogen and Deuterium
X-Rays } \\
\vspace*{0.2cm}
A.~Badertscher,
{\em Institute for Particle Physics, ETH Zurich, Switzerland }
\end{center}

The 3p-1s x-ray transitions in pionic hydrogen and deuterium were measured with
a high-resolution reflecting crystal spectrometer. The energy level shifts
$\varepsilon_{1s}$ and decay widths $\Gamma_{1s}$ of the 1s state,
induced by the strong interaction, lead to a determination of the two
$\pi$N s-wave scattering lengths directly at threshold [1,2].
Preliminary results for pionic hydrogen are [3]:
$\varepsilon_{1s} = -7.108 \pm 0.013(stat.) \pm 0.034(syst.)~eV$ (attractive)
and $\Gamma_{1s} = 0.897 \pm 0.045(stat.) \pm 0.037(syst.)~eV$.
Inserting these values into Deser's formula [4], and adding the
statistical and systematic errors linearly, results in the following
preliminary values for the scattering lengths for elastic and charge
exchange scattering: $a_{\pi^-p\rightarrow\pi^-p}^h = (0.0883
\pm 0.0008)~m^{-1}_{\pi}$ and $a_{\pi^-p\rightarrow\pi^{\circ}n}^h =
(-0.1301 \pm 0.0059)~m^{-1}_{\pi}$. \\
The systematic error of the decay width $\Gamma_{1s}$
is due to the uncertainty of the Doppler correction of the
measured line width. This Doppler broadening is due to a gain in kinetic
energy of the pionic atoms after a Coulomb-transition occured during the
cascade [5]. To study the kinetic energy distribution of the pionic
atoms, a new experiment was performed at PSI, measuring the time of flight of
neutrons emitted after the charge exchange reaction
$\pi^- p \rightarrow \pi^{\circ} n$. A
first measurement [6], with pionic atoms formed in liquid hydrogen,
confirmed the result of ref. [7], that about half of the pionic atoms have
kinetic energies $\gg$ 1~eV (up to $\simeq$ 200~eV). This summer,
a measurement
with gaseous hydrogen (density approx. 40 $\rho_{STP}$) was made, since the
pionic x-rays were also measured from gaseous hydrogen or deuterium.
The result will be used for a new calculation
of the Doppler broadening, yielding a final value for the decay width of the
1s state of pionic hydrogen and the charge exchange scattering length. \\
The (final) results from pionic deuterium are [2]:
$\varepsilon_{1s} = + 2.43 \pm 0.10~eV$ (repulsive) and
$\Gamma_{1s} = 1.02 \pm 0.21~eV$, yielding the complex $\pi^-$d scattering
length: $a_{\pi^-d} = (-0.0259 \pm 0.0011) + i(0.0054 \pm 0.0011)~m^{-1}_{\pi}$.
The real part of the $\pi^-$d scattering length can be related to the $\pi$N
scattering lengths [8,9,10]. \\

\vspace*{0.2cm}

\noindent [1] D.~Sigg et al., Phys.\ Rev.\ Lett.\ {\bf 75}, 3245 (1995), \\
D.~Sigg et al., Nucl.\ Phys.\ {\bf A609}, 269 (1996).

\noindent [2] D.~Chatellard et al., Phys.\ Rev.\ Lett.\ {\bf 74}, 4157 (1995),
\\
D.~Chatellard et al., to be published in Nucl.\ Phys.\ A.

\noindent [3] H.-Ch.~Schr\"oder, PhD thesis No.11760, ETH Zurich, 1996,
unpublished.

\noindent [4] S.~Deser et al., Phys.\ Rev.\ {\bf 96}, 774 (1954), \\
D.~Sigg et al., Nucl.\ Phys.\ {\bf A609}, 310 (1996).

\noindent [5] E.C.~Aschenauer et al., Phys.\ Rev.\ {\bf A 51}, 1965 (1995).

\noindent [6] A.~Badertscher et al., Phys.\ Lett.\ {\bf B 392}, 278 (1997).

\noindent [7] J.F.~Crawford et al., Phys.\ Rev.\ {\bf D 43}, 46 (1991).

\noindent [8] A.W.~Thomas and R.H.~Landau, Phys.\ Rep.\ {\bf 58}, 121 (1980).

\noindent [9] V.V.~Baru and A.E.~Kudryavtsev, $\pi$N Newsletter {\bf No.12}, 64
March  1997.

\noindent [10] S.R. Beane et al., preprint nucl-th/9708035.

\pagebreak

\noindent
{\large\bf Status of  $\sigma$--term calculations}

\bigskip

\noindent Bu\=gra Borasoy\\
Department of Physics and Astronomy,
{University of Massachusetts},
{Amherst, MA 01003}, USA

\bigskip

\noindent The $\sigma$--terms are defined by
\begin{eqnarray}
\sigma_{\pi N} (t) & = & \hat m \, <p' \, | \bar u u + \bar d d| \, p>
\, \, \, , \nonumber \\
\sigma_{KN}^{(1)} (t) & = & \frac{1}{2}(\hat m + m_s) \, 
<p' \, | \bar u u + \bar s s| \, p> \, \, \, , \nonumber \\
\sigma_{KN}^{(2)} (t) & = & \frac{1}{2}(\hat m + m_s) \, 
<p' \, | -\bar u u + 2\bar d d + \bar s s| \, p> \, \, \, , \nonumber 
\end{eqnarray}
with $| \, p>$ a proton state with four--momentum $p$, 
$t = (p'-p)^2$ the invariant momentum transfer squared and
$\hat m = ( m_u + m_d )/2$ the average light quark mass.
In this talk I discussed some of the results presented in [1].
This was the first calculation including all terms of second order in 
the quark masses (fourth order in the meson masses). The calculations were performed 
in the isospin limit $ m_u = m_d $ and the electromagnetic corrections were 
neglected. The most general effective Lagrangian to fourth order necessary to 
investigate the 

$\sigma$--terms consists of fourteen unknown coupling 
constants (LECs). Since we are not able to fix them from data we estimate them from 
resonance exchange. It turns out that for the scalar--isoscalar LECs one has to 
consider besides the standard tree graphs with scalar meson exchange also Goldstone 
boson l

oops with intermediate baryon resonances (spin--3/2 decuplet and 
spin--1/2 (Roper) octet). To leading order in the resonance masses the pertinent 
graphs are divergent. Using the baryon masses and $\sigma_{\pi N} (0)$
as input one can determine the a priori unknown renormalization constants.
The chiral expansion of the $\pi N$ $\sigma$--term shows a 
moderate convergence :
$\, 
 \sigma_{\pi N} (0) = 58.3 \, ( 1 - 0.56 + 0.33) \, \mbox{MeV} = 45 
\, \mbox{MeV} \, $.
The strangeness fraction $y$ and $ \hat \sigma$ are defined via
\begin{eqnarray}
y = \frac{2 <p \, | \bar s s| \, p>}{<p \, | \bar u u + \bar d d| \, p>}
= 1 - \frac{\hat \sigma}{\sigma_{\pi N} (0)} \qquad . \nonumber
\end{eqnarray}
We obtain
$ \, y = 0.21 \pm 0.20 \,\,\mbox{and} \, \, 
\hat \sigma = (36 \pm 7 )\, \mbox{MeV}
\, . $
In the case of the $KN$ $\sigma$--terms the results can only be given up to two 
renormalization constants which are an artifact of a calculation with 
$ m_u = m_d $. Varying these constants between 0.5 and 1 leads to 
$ \,
\sigma_{K N}^{(1)} (0) = 73 \ldots 216 \, \mbox{MeV} \, \, \mbox{and}  \, \, 
\sigma_{K N}^{(2)} (0) = 493 \ldots 703 \, \mbox{MeV} \, . $
These numbers are only indicative and have to be sharpened in a calculation with $ 
m_u \neq m_d $.
The shifts to the pertinent Cheng--Dashen points are
$ \,
\sigma_{\pi N} (2 M_{\pi}^2) - \sigma_{\pi N} (0)= 5.1 \, \mbox{MeV}
\, ,\,
\sigma_{K N}^{(1)} (2 M_{K}^2) - \sigma_{K N}^{(1)} (0)= (292 + i \, 365) 
\, \mbox{MeV} \, \, \mbox{and} \, \,  
\sigma_{K N}^{(2)} (2 M_{K}^2) - \sigma_{K N}^{(2)} (0)= (- 52 + i \, 365) 
\, \mbox{MeV} \,$ .

\bigskip

\noindent
[1] B. Borasoy and Ulf--G. Mei{\ss}ner, Ann. Phys. {\bf 254} (1997) 192

\pagebreak

\noindent
{\Large \bf The sigma-term revisited} 

\bigskip

\noindent
M.E. Sainio  \\
Dept. of Physics, Univ. of Helsinki, P.O. Box 9, FIN-00014 Helsinki, Finland

\bigskip

\noindent The pion-nucleon $\Sigma$-term, which is essentially the isoscalar
$D$-amplitude at the Cheng-Dashen point with the pseudovector Born term
subtracted, is a sensitive quantity. The widely accepted values have been
in the range 60-65 MeV with an uncertainty of about 10-12 MeV. However,
the discussions in the MENU97 meeting in Vancouver have confused the situation 
considerably. The results presented there, based on a whole variety of
approaches, may be summarized by $\Sigma$ = $60\pm20$ MeV, if
all the proposed values are to be included.

The dispersion method discussed in Ref. [1] and applied in [2] is
appropriate for determining the low-energy amplitudes and for extrapolating
to the near-by unphysical region. In particular, the aim in [2] was to
estimate the effect of the experimental errors of the low-energy data to
the uncertainty of the $D^+$-amplitude at the Cheng-Dashen point.
The method, however, needs as input amplitudes which satisfy fixed-$t$
dispersion relations with good precision. Beyond the Karlsruhe amplitudes
the VPI group has recently started to incorporate analyticity
constraints into their partial wave analysis [3]. This is important,
because the Karlsruhe amplitudes are based on data which were available
about 1980, and quite a few results have been published since then. The
new data also revise our understanding especially at the lowest energies.

In the extrapolation from $t=0$ to the Cheng-Dashen point, $t=2 \mu^2$,
the three contributions to the $\Sigma$, the constant, the linear 
part and the curvature contribution are 
about $-$90 MeV, 140 MeV and 12 MeV respectively. The largest term, 140 MeV, 
is fixed by the forward dispersion relation for the $E^+$-amplitude [1].
The partial wave expansion for the $E$-amplitude contains high powers of
angular momentum, $l^3$,  which induces some sensitivity to the
$d$-waves even at low energy. In Ref. [2] the error estimate contained a
$\pm 30 \%$ uncertainty for all the $d$-waves which at that time seemed
very generous. The low-energy results are not sensitive to the differences
at higher energies which can be tested by comparing the results of the 
KH.80 and KA.84 input amplitudes. The main
change is at the low energies where the observable cross sections are
evaluated and compared with the experimental results. The main parameters
in the fit are the two subtraction constants, isoscalar $s$-wave scattering
length and the upper $p$-wave scattering volume, which are fixed with
the constraint that the forward dispersion relations are exactly satisfied.
 
\noindent [1] J. Gasser et al., Phys. Lett. B {\bf 213} 85 (1988) 

\noindent [2] J. Gasser, H. Leutwyler and M.E. Sainio, Phys. Lett. B {\bf 253} 
252 (1991)

\noindent [3]  M. Pavan, these Proceedings

\pagebreak

\begin{center}

{\Large \bf V.P.I. $\pi$N PWA: Recent results for $\Sigma$ and f$^{2}$}

Marcello\,M.\,Pavan

Lab for Nuclear Science, M.I.T., 77 Massachusetts Ave., Cambridge,
  MA 02139
\end{center}

The V.P.I. partial--wave analysis group updates its pion--nucleon
partial wave analysis (PWA) regularly [1] as high quality data, 
 much still at low and $\Delta$ resonance energies, emerges steadily from the
 world's meson factories.  Updating is 
necessary as the Delta resonance region is crucial to 
phenomenological extractions of the $\pi$N Sigma $\Sigma$ term and $\pi$NN
coupling constant f$^{2}$.

It is well known that analyticity constraints are essential in PWA to reliably
extract $\Sigma$ and f$^{2}$ from scattering data.  The V.P.I. analysis
implements constraints from forward C and E (t-derivative) dispersion
relations (DR), and fixed-t dispersion relations (0 to t=-0.3 GeV$^{2}$) for
the invariant B and A amplitudes, from threshold to $\sim$700 MeV.  The DRs
are constrained to be satisfied to within $\sim$1\% over the relevant
kinematical ranges. The {\it a priori} unknown constants of the forward DRs
and B fixed--t DR, the s--wave (p--wave) a$_{0}^{\pm}$ (a$_{1+}^{+}$)
scattering lengths (volume), and the coupling constant f$^{2}$, are treated as
parameters to be determined by least squares fitting of the data and the DRs
together. Recent solutions indicate a best fit for f$^{2}$=0.0760$\pm$0.0005,
a$_{0}^{+}\sim$0.0004, a$_{\pi^{-}p}\sim$0.088 m$^{-1}_{\pi}$, and
a$_{1+}^{+}\sim$0.136 m$^{-3}_{\pi}$. The scattering length results agree
with the results derived from the recent precise line shift and width
measurements on hydrogen and deuterium [2].

Using the method of Ref. [3], and also by
extrapolating the  A DR subtraction constants (A$^{+}$(0,t)) linearly to the
Cheng-Dashen point, we find a Sigma
term $\Sigma_{d} \sim 75 MeV$, compared to the "canonical" result $\sim$50 MeV
[3]. It must be stressed that there is little 
reason beyond nostalgia to adhere to the "canonical" result, which uses 
a PWA solution [4] based on older (pre-meson factory), sparse, and often
outdated data.  The current VPI solution 
satisfies much better than [4] the relevant fixed-t and forward DRs, and also
provides a much better fit to the world data (and almost any subset),
especially around the Delta resonance. {\it e.g.} the fit to the $<$350 MeV
$\pi^{-}$p polarization data is 3x worse using  solution [4], and it is
shown in [5] that P($\pi^{-}$p) is sensitive to the constant c$^{+}_{00}$,
which contributes 1/3 to $\Sigma_{d}$.  A {\it new} PWA
analysis based {\it e.g.} on the methods in [4] is called upon to check our
results for $\Sigma$ and f$^{2}$.

%
%

\noindent [1] V.P.I. $\pi$N PWA solutions : TELNET
  claid.phys.vt.edu  (username : 'said') 

\noindent [2] A. Badertscher, these proceedings

\noindent [3]
 J. Gasser, H. Leutwyler and M.E. Sainio (1991) :
Phys.\,Lett.\,B{\bf 253} 252

\noindent [4]
R. Koch, E. Pietarinen (1980) :
Nucl.\,Phys.\, A{\bf 336} 331 

\noindent [5]
R. Koch (July 1985) :
Karlsruhe preprint TKP 85-5

\pagebreak

\bigskip
\noindent
\begin{center}
\noindent
{\Large \bf Dual structure of Chiral expansions.} 

\underline{Vladimir V.Vereshagin},
Alexander V.Vereshagin

Institute of physics, St.-Petersburg State University

St.-Petersburg, 198904, Russia.
\end{center}
\bigskip

\noindent                           

Two main problems preventing the further successful application
of the methods of Chiral Perturbation Theory are the following:
1. The problem of low energy constants (LEC's);
2. The problem of resonances. They are closely connected with each
other and the property of Chiral Duality (or, the same, resonance
saturation --- see, e.g. [1]) clearly demonstrates this connection.
Unfortunately, Chiral Duality tells us nothing about the structure
of LEC's in exotic channels and, of course, it does not provide
a solution to the problem of resonances.
Below we give a sketch of the general approach allowing one to solve
both problems simultaneously. The main idea of the approach is to take
advantage of the old good principle of maximal analiticity, the
latter being understood as the requirement of meromorphy and polynomial
boundedness of the tree level amplitude considered as a function of
3 dependent complex variables
$ s,\;t,\;u$.
This requirement proves to be highly nontrivial when applied to
\underline{effective}
field theory containing
\underline{all}
possible types of the interaction vertices.

Let us now write down two most important (and constructive) postulates.
1.The tree-level amplitude of every binary process is a meromorphic
function in a space of 3 dependent complex variables
$ s,\; t,\; u$
with the principle part completely fixed by the corresponding ``naive''
Feynman rules.
2. When considered as a function of any
\underline{one}
complex variable
$x$
(the energy squared in a given channel) at fixed real nonpositive value
$y$
of the corresponding momentum transfer the amplitude is polynomially
bounded in
$x$,
the degree of bounding polynomial being dictated by the relevant Regge
intercept.
The main tool for derivation of results is provided by the Mittag-Leffler
theorem.
This allows us to write down three different partial fraction expansions
for the amplitude, each one being convergent in a corresponding band
adjacent to one of the sides of Mandelstam triangle. The self-consistency
requirements in the domains of mutual intersection of those bands turn out
to be just the duality conditions which produce an infinite set of
bootstrap equations for the resonance masses and widths. These latter
equations allow one to express the LEC's in terms of spectrum parameters.

The fruitfulness of our approach is illustrated in detail in
[2]
by the example of
$\pi K$ - amplitude; its connection with Weinberg's result
on algebraic realization of Chiral symmetry is considered in
[3].\\
\\

\noindent [1] Ulf-G.Meissner. Rep. Progr. Phys., \underline{56} (1993) 903.

\noindent [2] V.Vereshagin. Phys. Rev., \underline{D55} (1997) 5349.

\noindent [3] V.Vereshagin, M.Dillig, A.Vereshagin. HEP-PH/9611392;
Preprint SPbU-IP-96-33 (1996).

\end{document}